\documentclass[aps,pra,twocolumn,showpacs]{revtex4-1}    

\usepackage{amsmath}
\usepackage{amsfonts}
\usepackage{amssymb}
\usepackage{amscd}
\usepackage{amsbsy}         
\usepackage{bm}             
\usepackage{graphicx}       
\usepackage{dcolumn}        
\usepackage{latexsym}
\usepackage[latin1]{inputenc}   
\usepackage{float}          

\newcommand  {\lat}   {\mathrm{lat}}
\newcommand  {\eff}   {\mathrm{eff}}

\newcommand  {\kB}    {k_\mathrm{B}}
\newcommand  {\e}     {$|e\rangle$}
\newcommand  {\g}     {$|g\rangle$}
\newcommand  {\m}     {$|m\rangle$}
\newcommand  {\ii}    {$|i\rangle$}
\newcommand  {\jj}    {$|j\rangle$}

\newcommand {\tub} {Physikalisches Institut, Eberhard-Karls-Universit\"{a}t T\"{u}bingen, Auf der Morgenstelle 14, D-72076 T\"{u}bingen, Germany}

\begin{document}

\title{Photonic properties of one-dimensionally-ordered cold atomic vapors under conditions of electromagnetically induced transparency}

\author{Alexander Schilke}
\author{Claus Zimmermann}
\author{William Guerin}
\email{william.guerin@pit.uni-tuebingen.de}
\affiliation{\tub}
\date{\today}

\begin{abstract}
{We experimentally study the photonic properties of a cold-atom sample trapped in a one-dimensional
optical lattice under the conditions of electromagnetically induced transparency. We show that such a medium has two photonic band gaps. One of them is in the transparency window and gives rise to a Bragg mirror, which is spectrally very narrow and dynamically tunable. We discuss the advantages and the limitations of this system. As an illustration of a possible application we demonstrate a two-port all-optical switch.}
\end{abstract}


\maketitle

\section{Introduction}   

Atomic vapors can be used for studying many original or useful optical phenomena. Based on the atomic nonlinearity, one can produce and study bistability \cite{Rempe:1991}, squeezing \cite{Lambrecht:1996}, various nonlinear magneto-optical effects \cite{Budker:2002}, all-optical switching \cite{Dawes:2005}, gain and lasing \cite{LasingWithColdAtoms}, and four-wave mixing \cite{Bloom:1978}, which allows the production of twin beams \cite{Vallet:1990,Boyer:2008b} and optical parametric oscillation \cite{Pinard:1986}.
Another useful property is the atomic coherence, that can be used to produce electromagnetically induced transparency (EIT) \cite{Harris:1997,Fleischhauer:2005}, slow or fast light \cite{Hau:1999,Akulshin:2010} and quantum memories \cite{Sangouard:2011}.
Finally, the large atomic scattering cross section allows studying effects related to multiple scattering of light in disordered media, for example Lévy flights in hot vapors \cite{Mercadier:2009}, radiation trapping \cite{Fioretti:1998,Labeyrie:2003}, and coherent backscattering in cold atoms \cite{Labeyrie:1999,Bidel:2002}. In the opposite regime, cold atoms can be trapped in an ordered fashion, which gives rise to Bragg scattering \cite{Birkl:1995,Weidemuller:1995,Slama:2005a} and photonic band gaps (PBGs), which have been recently observed in the one-dimensional (1D) case \cite{Schilke:2011} and predicted in three dimensions \cite{Antezza:2009,Yu:2011}.

Combining a control over the atomic spatial arrangement (external degrees of freedom) and the atom polarizability (internal degrees of freedom) allows a complex engineering of the propagation properties of light. In this spirit, radiation trapping under condition of EIT has been studied in \cite{Matsko:2001,Datsyuk:2006}, the combination of multiple scattering and gain gives rise to random lasing \cite{RandomLaser}, and it was recently demonstrated that the combination of a 1D PBG with four-wave mixing leads to distributed feedback optical parametric oscillation \cite{Schilke:2012}.

In this paper, we experimentally investigate the combination of EIT and a 1D PBG formed by cold atoms trapped in a 1D lattice, like in \cite{Schilke:2011}. As already shown in a theoretical paper by Petrosyan \cite{Petrosyan:2007}, such a system creates a new band gap, in the transparency window, which is spectrally very narrow and which is dynamically tunable. We report measurements of the transmission and reflection spectra and their dependence with experimental parameters, and we discuss the limitations of such a system. We finally demonstrate a two-port all-optical switch as a possible application.

It should be noted that various configurations of electromagnetically induced gratings have already been discussed in the literature with hot or cold atoms (see \cite{EITgrating_theory} for theoretical proposals and \cite{EITgrating_exp} for experiments). In all these cases, however, the grating is due to the spatial modulation of the control field. On the contrary, in our experiment, the Bragg mirror relies on the periodic spatial modulation of the atomic density. The controls over the internal and external degrees of freedom are thus decoupled.

The paper is organized as follows. The next section is devoted to the theoretical description of our system. The dispersion relations and expected reflection and transmission spectra are computed for ideal parameters. In the following part, we present our experimental setup. Then, in Sec. \ref{sec.meas}, we present our measurements. Finally, in Sec. \ref{sec.switch}, we demonstrate the use of our system as an all-optical switch.

\section{Theoretical description} \label{sec.theory} 

\begin{figure}[t]
\centering \includegraphics{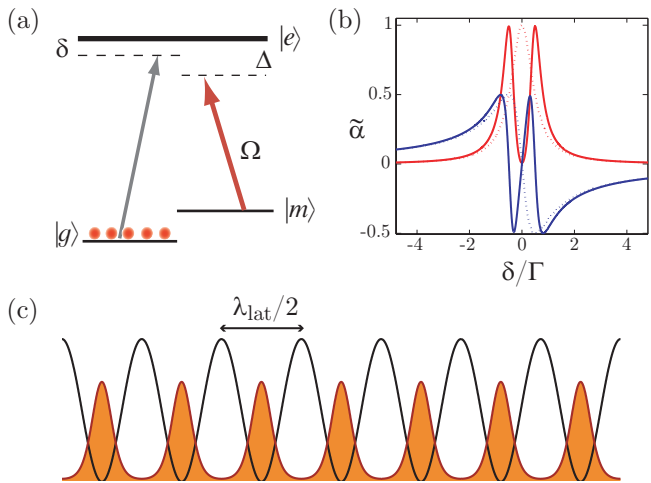}
\caption{(Color online) (a) Atomic levels and laser configuration. (b) Real part (blue, bottom curve) and imaginary part (red) of the dimensionless atomic polarizability $\tilde{\alpha}=\alpha\times \varepsilon_0\hbar\Gamma /2|d_{ge}|^2$ for a two-level atom (dotted lines) and with EIT (solid lines), with $\Delta = 0$ and $\Omega = \Gamma$ [Eq.~(\ref{eq.alpha})]. (c) Scheme of the system under consideration: the atoms are trapped in a 1D optical lattice of periodicity $\lambda_\lat/2$.}
\label{fig.theory}
\end{figure}

We consider three-level atoms, as shown in Fig.~\ref{fig.theory}(a), with two low-energy levels, which we call the ground state \g\ and the metastable state \m, and one excited state \e. The atoms are initially in the ground state, and we are interested in the photonic response of the sample at optical frequencies in the vicinity of the transition \g $\leftrightarrow$\e (wavelength $\lambda_0$), when the states \m , \e\ are coupled by an external field. We thus consider a probe beam with a detuning $\delta = \omega - \omega_{ge}$ from the atomic transition and a coupling beam with a detuning $\Delta = \omega_\mathrm{C} - \omega_{me}$. The probe beam has a very low intensity and we consider only the atom's linear response, described by the atomic polarizability [Fig.~\ref{fig.theory}(b)]
\begin{equation}\label{eq.alpha}
\alpha = \frac{2|d_{ge}|^2}{\varepsilon_0\hbar\Gamma} \times \frac{-\Gamma}{2\delta+i\Gamma-\Omega^2/[2(\delta-\Delta+i\gamma)]} \; ,
\end{equation}
where $\Gamma$ is the spontaneous emission rate of the excited state, $\gamma$ the dephasing rate between the two ground states (we suppose $\gamma<<\Gamma$), $\Omega = |d_{me}|^2E/\hbar$ is the Rabi frequency of the coupling field of amplitude $E$, and $d_{ij}$ is the dipole moment of the transition \ii\ $\leftrightarrow$ \jj\ \cite{Petrosyan:2007}. In this equation, EIT is induced by the last term of the denominator \cite{footnote_alpha}.

The atoms are trapped in a one-dimensional optical lattice formed by a red-detuned retroreflected laser beam of wavelength $\lambda_\lat$, thus forming an atomic density grating of periodicity $\lambda_\lat/2$ [Fig.~\ref{fig.theory}(c)]. The modulation contrast depends on the temperature $T$ of the atomic sample, which is usually related to the trapping depth $U_0$ of the optical potential by a constant factor $\eta = U_0/\kB T$. We will take $\eta = 3.5$, the value observed in our experiment \cite{Schilke:2011}. The density distribution of each period is then a Gaussian of rms width along the lattice axis $z$ $\sigma_z = \lambda_\lat/(2\pi\sqrt{2\eta})$.

Since the laser forming the lattice must have a wavelength $\lambda_\lat > \lambda_0$ to create a dipole trap, the Bragg condition can only be fulfilled with a nonzero propagation angle $\theta$ between the probe and the lattice beams, such that $\cos\theta \sim \lambda_0/\lambda_\lat$. In practice, it is easier in experiments to tune the lattice wavelength to adjust the Bragg condition. We can thus define $\Delta\lambda_\lat = \lambda_\lat- \lambda_{\lat0}$ as the shift from the ``geometric" Bragg condition $\lambda_{\lat0} = \lambda_0/\cos \theta$. The complete Bragg condition must take into account the fact that the probe wavelength in the medium is $\lambda = \lambda_0/n$, where $n$ is the average refractive index, which strongly depends on the probe detuning $\delta$. The Bragg condition can then be rewritten in the simple following form,
\begin{equation}\label{eq.bragg}
n(\delta)-1 = -\frac{\Delta\lambda_\lat}{\lambda_{\lat}}  ,
\end{equation}
where the right-hand side of the equation depends only on the lattice wavelength and the left-hand side depends on the real part of the atomic polarizability and on the average atomic density $\rho$, with $n-1 = \rho/2\times \mathrm{Re}(\alpha)$ for a dilute vapor. The imaginary part of the atomic polarizability plays also an important role since it is responsible for scattering losses \cite{Schilke:2011}.

Considering these losses together with Eq.~(\ref{eq.bragg}) is sufficient to qualitatively explain the photonic properties of the system (see \cite{Schilke:2011} for the simple case of two-level atoms): a band gap will appear when Eq.~(\ref{eq.bragg}) is fulfilled at a detuning $\delta$ where the imaginary part of the atomic polarizability is small enough. With the EIT polarizability [Eq.~(\ref{eq.alpha}) and Fig.~\ref{fig.theory}(b)], one can easily see that the Bragg condition (\ref{eq.bragg}) can be fulfilled at four different detuning $\delta$ (crossing points between Re$(\alpha)$ and a straight horizontal line given by $-\Delta\lambda_\lat/\lambda_{\lat}$). However, for two of these frequencies, the imaginary part of the polarizability is near its maximum, indicating a large amount of losses and preventing any efficient Bragg reflection. Therefore, we expect two band gaps, one for a large detuning, which also appears with two-level atoms \cite{Schilke:2011}, and another one, narrower, in the transparency window, which is due to EIT.

A more precise description of the photonic properties of such a periodic atomic structure can efficiently be obtained by simulating light propagation in the medium with the transfer matrix method \cite{BornWolf,Lekner:1994,Bendickson:1996}. It is a one-dimensional model, whose use is justified when the transverse extension of the atomic layers is large compared to the probe beam size and when the incident angle is small, which is the case in our experiment (see \cite{Slama:2005b} for an extended discussion on this issue). The nonzero propagation angle can be taken into account by changing the probe wavevector from $k_0=2\pi/\lambda_0$ to $k_0 \cos\theta$.
A detailed description of this method in the context of ordered atomic samples has been given in previous papers \cite{Deutsch:1995,Artoni:2005,Slama:2006}. In brief, the first step is to construct the transfer matrix $M$ of one single period. To do so, one has to decompose the atomic layer in several sublayers of thickness $\delta z$. The transfer matrix of each sublayer is the product of a propagation matrix with a discontinuity matrix whose coefficients are given by the Fresnel coefficients, see \cite{Bendickson:1996}. Besides the density distribution, the only ingredient entering the model is the atomic polarizability. Therefore, extending the results obtained with two-level atoms \cite{Schilke:2011} to driven atom under EIT conditions is simply made by replacing the atomic polarizability.
Once the matrix $M$ is obtained, we can use it to derive analytical formula that allow us to compute the dispersion relations and the reflection and transmission coefficients through $N$ layers (see e.g. \cite{Lekner:1994,Deutsch:1995,Bendickson:1996}). The matrix $M$ is related to the elementary reflection $r$ and transmission $t$ coefficients of one single period by
\begin{equation}\label{eq.M}
M = \frac{1}{t} \begin{bmatrix} t^2-r^2 & r \\ -r & 1 \end{bmatrix} .
\end{equation}
Then, using the property $\det(M) = 1$, the eigenvalues of $M$ are $e^{\pm i \Theta}$ with
\begin{equation} \label{eq.Theta}
\cos(\Theta) = \cos(k_\eff \frac{\lambda_\lat}{2})= \frac{\mathrm{Tr}(M)}{2}.
\end{equation}
This relation gives the effective wavevector (or Bloch wavevector) $k_\eff$ in the medium, i.e. the dispersion relation, which describes the photonic properties of the medium in the limit where it is infinite.

To compute the transmission and reflection coefficients through $N$ periods, we first introduce the matrix $A$ such that
\begin{equation}\label{eq.MA}
M = e^{i\Theta A} = \cos(\Theta) I + i \sin(\Theta) A ,
\end{equation}
where $I$ is the $2\times2$ identity matrix.
Then, the transfer matrix of $N$ periods writes
\begin{equation}\label{eq.MN}
M^N = e^{iN\Theta A} = \cos(N\Theta) I + i \sin(N\Theta) A .
\end{equation}
To get the transmission coefficient $T=|t_N|^2$ and reflection coefficient $R=|r_N|^2$  from $M^N$, we just need to know the coefficients of $A$, which are obtained by Eq. (\ref{eq.MA}) and identification with  Eq. (\ref{eq.M}). After some algebra, we get
\begin{eqnarray}
r_N & = & \frac{r}{1-t [\cos(\Theta) - \sin(\Theta)\cot(N\Theta)]} \; , \label{eq.rN}\\
t_N & = & \frac{t \sin(\Theta)/\sin(N\Theta)}{1-t [\cos(\Theta) - \sin(\Theta)\cot(N\Theta)]} \; . \label{eq.tN}
\end{eqnarray}
For an infinite medium, and with Im$(\Theta)>0$, we obtain
\begin{equation}\label{eq.r_inf}
r_\infty  =  \frac{r}{1-t e^{i\Theta}} .
\end{equation}

\begin{figure}[t]
\centering \includegraphics{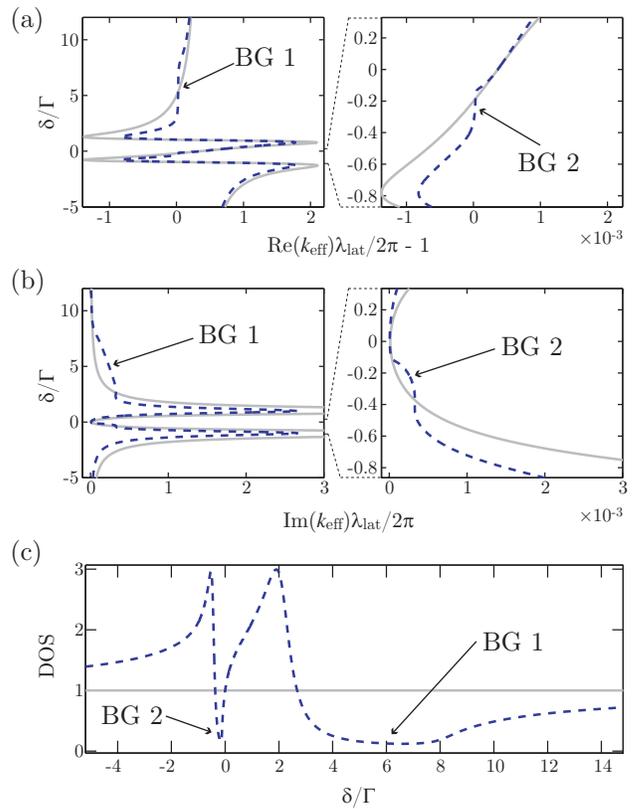} \caption{(Color online) Photonic properties of the medium, valid for an infinitely long lattice. (a) Dispersion relation: the frequency (detuning $\delta$) is plotted as a function of the effective wavevector in the medium Re$(k_\eff)$. Only the edge of the first Brillouin zone $k_\eff \lambda_\lat/2 =\pi$ is shown. A first band gap (BG) is visible. Right panel: zoom in the transparency window, where a second BG appears. (b) Same as (a) with the imaginary part of $k_\eff$. (c) Density of states (DOS) normalized to the one in a bulk medium of the same susceptibility [Eq.~(\ref{eq.DOS})]. In all plots, the gray solid lines correspond to a homogeneous atomic medium of the same
average density $\rho = 7\times10^{11}$~cm$^{-3}$ and the dashed blue lines correspond to atoms trapped in a lattice with $\eta=3.5$, $\Delta\lambda_\lat = 0.25$~nm and coupling-field parameters $\Omega=2\Gamma$ and $\Delta=0$.}
\label{fig.Dispersion}
\end{figure}

We applied these results with the $F=1 \rightarrow F'=2$ transition of the $D_2$ line of rubidium 87 and with the optimum parameters of \cite{Schilke:2011} ($\rho = 7\times10^{11}$~cm$^{-3}$, $\eta = 3.5$ \cite{footnote_sampling1} and $\Delta\lambda_\lat = 0.25$~nm) and with the coupling-beam parameters $\Delta=0$, $\Omega=2\Gamma$. We introduce a dephasing rate $\gamma = 0.008 \Gamma$, similar to the one of the experiment \cite{Hau:1999}. We obtain the dispersion relation [$\omega$ vs Re$(k_\eff)$] shown in Fig.~\ref{fig.Dispersion}(a). As expected from the previous qualitative discussion, it exhibits two band gaps (BGs), which appear at the edge of the first Brillouin zone $k_\eff \lambda_\lat/2 =\pi$, i.e. where the Bragg condition (\ref{eq.bragg}) is fulfilled. It is characterized by a reduced variation of Re$(k_\eff)$ with $\omega$, corresponding to a reduced density of states. One of the band gap, which we label ``BG 1'' in Fig.~\ref{fig.Dispersion}(a), is not influenced by EIT and is the same as the one studied in our previous experiment \cite{Schilke:2011}. The second one (``BG 2'') appears on the contrary in the electromagnetically induced transparency window. It is very much narrower and its width increases with the coupling-beam intensity.

The band gaps manifest themselves also in the imaginary part of $k_\eff$. In a lossless medium, $k_\eff$ acquires an imaginary part only in band gaps. In our system, since the atomic polarizability is complex, the wavevector has always an imaginary part leading to the wave attenuation when it propagates in the medium. This attenuation is due to scattering losses. In this case, we see in Fig.~\ref{fig.Dispersion}(b) that the BGs add an extra component of Im$(k_\eff)$, which is responsible for the formation of an evanescent wave that leads to the reflection of the incoming light.

Finally, PBGs appear also as a reduction of the density of states (DOS), which can be computed, following \cite{Boedecker:2003} and considering a position in the middle of the structure, from the reflection coefficients of the two surrounding semi-lattices of reflection coefficients $r_1$ and $r_2$, via
\begin{equation}\label{eq.DOS}
\mathcal{D} = \mathrm{Re}\left[\frac{2+r_1+r_2}{1-r_1 r_2} -1\right] .
\end{equation}
This can be applied for a finite length lattice in order to compute the local DOS \cite{Schilke:2011} or with an infinite lattice using Eq.~(\ref{eq.r_inf}). The result in that case is shown in Fig.~\ref{fig.Dispersion}(c) and demonstrates a strong DOS reduction in the two BGs. It should be noted that, despite the assumption of an infinite medium, the DOS does not completely vanish because of the scattering losses. It reaches a minimum value of $0.12$ (normalized to the DOS in the bulk medium of the same susceptibility).

\begin{figure}[t]
\centering \includegraphics{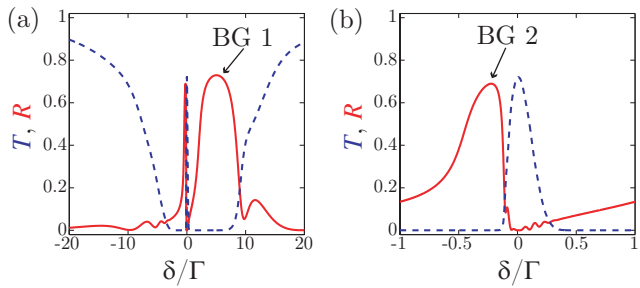}
\caption{(Color online) (a) Computed transmission $T$ (dashed blue line) and
reflection $R$ (red solid line) spectra with a lattice composed of $N\sim 7700$ periods. All other parameters are the same as for Fig.~\ref{fig.Dispersion}.} \label{fig.idealEIT}
\end{figure}

For a finite-size medium, the most relevant quantities are the transmission and reflection spectra, obtained from Eqs.~(\ref{eq.rN}) and (\ref{eq.tN}). They are shown in Fig.~\ref{fig.idealEIT}, where the two band gaps appear as two reflection bands. With the above-mentioned parameters and a lattice length $L=3$~mm, corresponding to $N\sim 7700$ periods if $\lambda_\lat \sim 781$~nm, the reflection coefficient reaches $R\sim 0.73$. Note that this is slightly lower than what is reported in \cite{Schilke:2011}, because the considered transition strength is weaker than the closed transition used in \cite{Schilke:2011}, which we cannot use for EIT.

To summarize, the use of EIT makes a new PBG appear in the transparency window, in addition to the one that already appears with two-level atoms. This result was already reported in \cite{Petrosyan:2007}. However, there is an important and natural question that has not been explicitly answered in \cite{Petrosyan:2007} (even if the result is visible in Fig. 4 of that paper): is the electromagnetically induced band gap (BG 2) of better quality than the other one (BG 1)? It was shown in \cite{Schilke:2011} that scattering losses were the main limitation for achieving low DOS or high reflectivity, and one could thus hope that EIT improves the band gap quality. As it can be seen in Fig.~\ref{fig.Dispersion}(c), where the DOS does not reach a lower value in the BG 2 than in the BG 1, and in Fig.~\ref{fig.idealEIT}, where the reflection coefficient is not higher in the BG 2 than in the BG 1, the answer to this question is that the EIT band gap is not of better quality. The explanation for this behavior is that even with a perfect EIT ($\gamma =0$), where complete transparency is reached (Im$(\alpha)=0$), it is reached precisely at a detuning $\delta$ where the real part of the atomic polarizability is also zero, thus suppressing any refractive-index grating. To build a PBG, one needs a nonzero refractive index, and the Bragg condition (\ref{eq.bragg}) can only be fulfilled slightly off the condition of perfect transparency. Moreover, the subsequent losses, given by Im$(\alpha)$, are exactly the same for both PBG. Taking into account an unperfect EIT ($\gamma >0$) leads even to slightly more losses.

Nevertheless, the EIT band gap has other advantages. The most important is that it is tunable, and dynamically controllable via the coupling-beam parameters. Moreover, it is very narrow and has a very sharp transition with a good transmission band [Fig.~\ref{fig.idealEIT}(b)]. These are interesting properties for practical applications, which motivate our experimental study, described in the following.

\section{Experimental setup} \label{sec.setup}  

In this section, we present briefly our experimental apparatus, which has already been described in \cite{Schilke:2011}.

\begin{figure}[t]
\centering \includegraphics{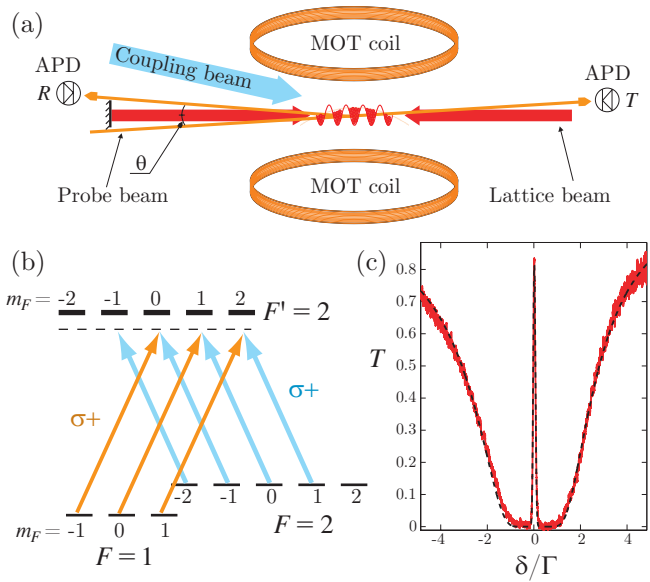}
\caption{(Color online) (a) Scheme of the experimental setup. (b) Atomic levels and laser configuration. The probe (orange) and coupling field (blue) drive the transition $F=1 \rightarrow F'=2$ and $F=2 \rightarrow F'=2$, respectively. (c) Transmission spectrum showing EIT with a disordered sample. The dashed black line is a fit to the data [Eqs.~(\ref{eq.alpha}) and (\ref{eq.trans})], yielding the parameters $\gamma = 7\times 10^{-3} \Gamma$, $\Omega = 0.8 \Gamma$ and optical thickness $b_0 = 21$. The slight asymmetry is due to the cloud expansion during the sweep and is taken into account in the fit.} \label{fig.setup}
\end{figure}

We trap and cool $^{87}$Rb in a magneto-optical trap (MOT) loaded from a background vapor. The optical lattice is generated by a homemade titanium-sapphire laser \cite{Zimmermann:1995} of maximum power $\sim 1.3$~W with a tunable wavelength $\lambda_\lat$. The beam is focused on a waist ($1/e^2$ radius) $w_\lat = 220\,\mu$m at the MOT position (Rayleigh length $z_\mathrm{R} \simeq 0.2$~m) and then retroreflected [Fig. ~\ref{fig.setup}(a)]. After stages of compression and molasses, the MOT is switched off and a waiting time of a few ms allows the untrapped atoms to fall down. The sample can then be characterized by absorption imaging or used for measuring transmission and reflection spectra. In this series of experiment, the typical atom number in the lattice is $N \sim 1-2 \times 10^7$ \cite{footnote_atomnumber}.

To acquire spectra, we shine a weak ($P \sim 3$~nW) and small (waist $w_0 = 35\,\mu$m) probe beam onto the lattice under an angle of incidence $\theta \simeq 1.5^\circ$, which is small enough to allow the beam to interact with the lattice over its entire length. The transmitted and reflected beams are then recorded with avalanche photodiodes (APDs). The probe detuning $\delta$ is swept in the vicinity of the atomic resonance by using an acousto-optical modulator in double-pass configuration. We use the $F=1 \rightarrow F'=2$ transition of the $D_2$ line ($\lambda_0 = 780.24$ nm, linewidth $\Gamma/2\pi=6.1$~MHz). The other hyperfine levels are far enough to be negligible. The presented data are the result of an average of typically 250 cycles (the duration of each cycle is $\sim 1$~s).

EIT is induced by a coupling beam tuned in the vicinity of the $F=2 \rightarrow F'=2$ transition [Fig.~\ref{fig.setup}(b)]. The beam has a diameter of about 5~mm and makes an angle with the lattice axis of about $8^\circ$, small enough to ensure a homogeneous coupling strength over the whole lattice. The probe and coupling beams have both the same circular polarization, which yields to complete EIT, since all Zeeman substates of the excited states are coupled to the metastable state \cite{Chen:2000,Yan:2001}. In addition, both lasers are phase-locked together via standard phase-locking techniques \cite{PLL} in order to fully exploit the coherence of the EIT process. To characterize the quality of the achieved EIT, we acquire a transmission spectrum with a disordered atomic sample by suddenly switching off the optical lattice and letting the atoms expand a few microseconds before sweeping the probe frequency in $200\,\mu$s. The ordered pattern has then disappeared and the transmission is given by
\begin{equation}\label{eq.trans}
T = \exp\left[-b_0 \mathrm{Im}(\tilde{\alpha})\right] \; ,
\end{equation}
where $b_0$ is the on-resonance optical thickness ($b_0 = \sigma_0 \int \rho(z) dz$ for a medium of density $\rho$ and with an on-resonance scattering cross section $\sigma_0$) and $\tilde{\alpha}$ is the dimensionless atomic polarizability, whose value is one at resonance (see its definition in the caption of Fig.~\ref{fig.theory}). Fitting a transmission spectrum by Eqs. (\ref{eq.trans}) with the polarizability (\ref{eq.alpha}) allows us to measure the on-resonance optical thickness, the effective dephasing rate $\gamma$ and to calibrate the Rabi frequency $\Omega$. With the recorded spectrum of Fig.~\ref{fig.setup}(c), we obtain $\gamma \sim 7\times 10^{-3} \Gamma$, giving for example a transmission of 81\% with an optical thickness $b_0 = 21$ and with only $\Omega = 0.8 \Gamma$ [Fig.~\ref{fig.setup}(c)]. The effective decoherence rate $\gamma$ mainly comes from the residual phase noise between the probe laser and the coupling lasers. Note also that the transparency increases with the coupling strength $\Omega$.

\section{Measurements} \label{sec.meas} 

\begin{figure}[t]
\centering \includegraphics{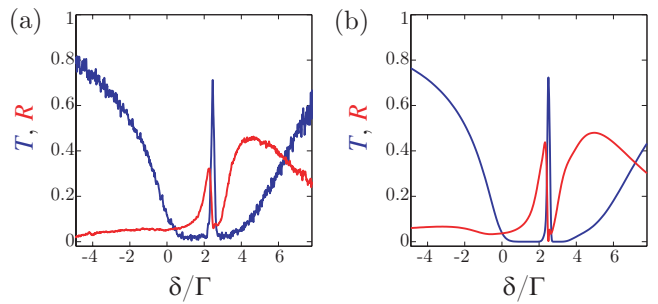}
\caption{(Color online) Experimental (a) and simulated (b) transmission (blue) and reflection (red) spectra, with $\Delta\lambda_\lat = 0.13$~nm, $N=1.5\times 10^7$ atoms, $\eta = 3.5$ and EIT parameters $\Omega = 1.3\Gamma$, $\Delta = 2.5\Gamma$, and $\gamma = 0.015 \Gamma$.} \label{fig.spectra}
\end{figure}

We now turn to our experimental characterization of the photonic properties of the cold-atom sample trapped in the lattice under EIT conditions. From now on, all spectra are taken with the lattice beam on. To begin with, let us examine an example of transmission and reflection spectra, shown in Fig.~\ref{fig.spectra}(a), recorded with $\Delta\lambda_\lat = 0.13$~nm and with the coupling-beam parameters $\Omega = 1.3 \Gamma$, $\Delta = 2.5\Gamma$. The coupling field is in fact almost resonant with the atomic transition, because the lattice trapping induces a light shift. This shift is slightly inhomogeneous because of the finite extension of the atomic cloud in each well, but at the potential minimum, where most atoms are, the light-shifted atomic resonance is at $\delta \sim 2.5 \Gamma$. This effect is taken into account in our simulations. Note that the light shift is the same for both transitions so that the two-photon resonance condition leading to EIT is not affected.

We clearly observe two reflection bands, as expected, corresponding to the two band gaps described in Sec. \ref{sec.theory}. The wide one is the band gap already studied in \cite{Schilke:2011}, while the narrow one, never observed before, appears in the transparency window and is due to EIT. We observe also that the reflection of the EIT band gap is lower than the reflection of the two-level-atom band gap. This is due to the finite dephasing rate $\gamma$, which explains also why the transparency is not complete in the transmission spectrum. However, taking into account this parameter in the simulation still leads to an overestimation of the reflection coefficient \cite{footnote_light_shift}. Apart from this discrepancy, whose origin remains unclear, the simulated spectra are in good agreement with the experimental ones [Fig.~\ref{fig.spectra}(b)].

\begin{figure*}
\centering \includegraphics{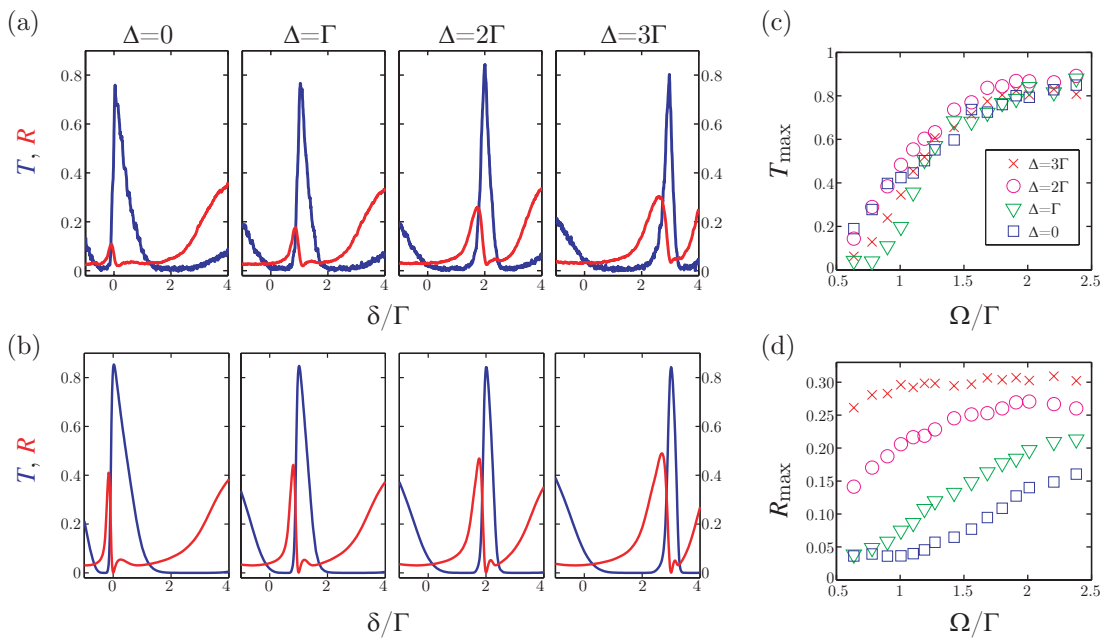}
\caption{(Color online) Dependency on the EIT parameters, with a fixed lattice wavelength $\Delta\lambda_\lat = 0.11$~nm. (a) Experimental transmission (blue) and reflection (red) spectra for several coupling-field detuning $\Delta$, with $\Omega = 1.8 \Gamma$. (b) Corresponding simulations. (c) Maximum transmission in the transparency window as a function of the coupling-field Rabi frequency $\Omega$, for different detunings $\Delta$. (d) Same as (c) with the maximum reflection in the EIT reflection band.} \label{fig.RT_vs_EIT}
\end{figure*}

The simulations shown here and in the following are more complicated than what has been described in Sec. \ref{sec.theory} because they take into account a number of experimental effects. Besides the above-mentioned light shift, the most important effect is the longitudinal atomic density distribution along the lattice, which is roughly Gaussian and can be precisely characterized by absorption imaging. This inhomogeneous distribution prevents the use of Eqs. (\ref{eq.rN}) and (\ref{eq.tN}). Instead, we have to compute a different elementary matrix for each position, following the measured atomic density distribution, and multiply them. Note that having a sample without sharp boundaries makes the usual dips and bumps at the band edges [Fig.~\ref{fig.idealEIT}(a)] disappear, inducing a kind of smoothing of the band edge. Another experimental effect that is included in the simulations is the inhomogeneous broadening due to the finite transverse sizes of the atomic sample and of the probe beam. The transfer-matrix formalism is a 1D model, but an approximate method to account for the transverse-size effects is to consider a distribution of probed densities corresponding to the overlap of the probe beam with the atomic lattice, and to average the subsequent spectra with the appropriate weighting. Since a finite probe size induces also some divergence and that the spectra are very sensitive to the incident angle, we average also over the corresponding angle distribution \cite{footnote_sampling2}. This procedure leads to a good agreement with the experiment (Fig.~\ref{fig.spectra}).

In the following, we will focus on the properties of the EIT reflection band and for clarity we will only show spectra in the corresponding, narrower spectral range. We will investigate the tunability of this reflection band, i.e., how it evolves with the coupling-beam parameters, and its dependence with the lattice wavelength via the Bragg condition (\ref{eq.bragg}). Both aspects are related because the Bragg condition involves the atom polarizability, which is modified by the EIT parameters. However, for simplicity, we separately present these two dependencies.

\subsection{Dependency on the EIT parameters}

To illustrate the tunability of the reflection and transmission bands, we show in Fig.~\ref{fig.RT_vs_EIT}(a) a series of spectra for different coupling-field detuning $\Delta$ with fixed intensity ($\Omega = 1.8 \Gamma$) and lattice wavelength ($\Delta\lambda_\lat = 0.11$~nm). As expected, the frequency giving the maximum transmission follows the two-photon resonance $\delta \simeq \Delta$, while the reflection band is slightly shifted on the $\delta<\Delta$ side. The corresponding simulations are in fair agreement with the experimental data, apart from the overestimated reflection in the EIT band gap [Fig.~\ref{fig.RT_vs_EIT}(b)].

\begin{figure*}[t]
\centering \includegraphics{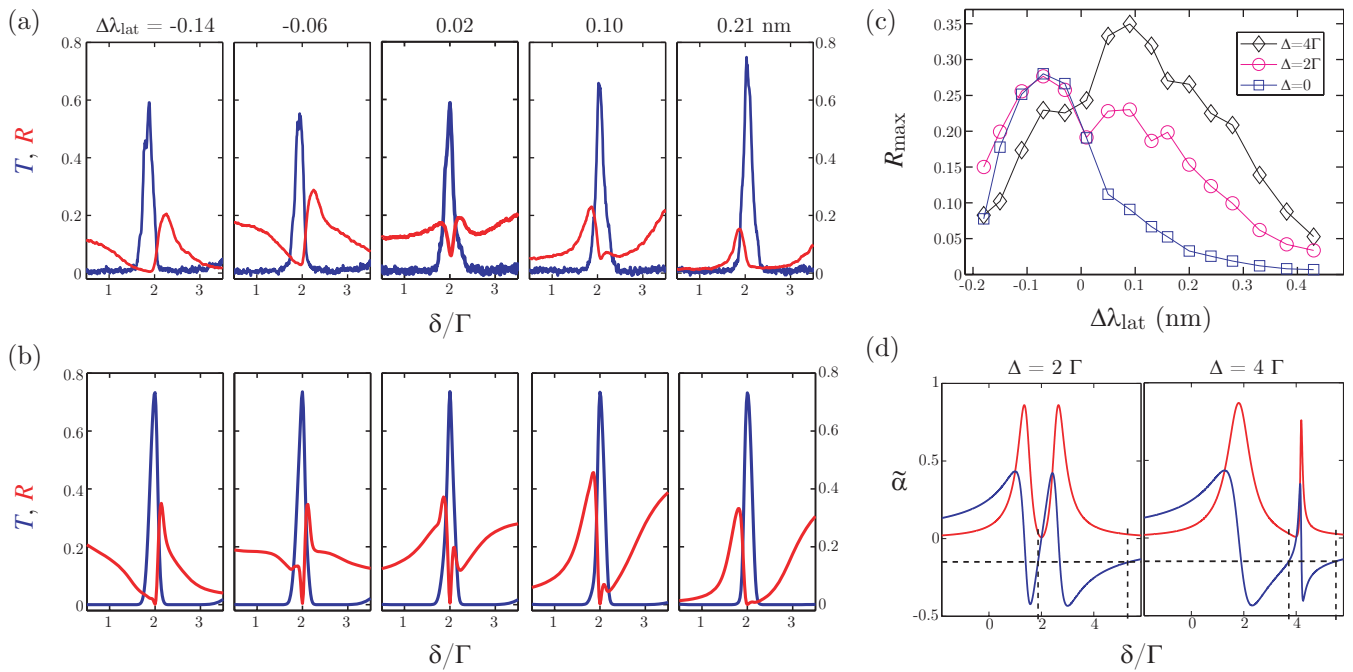}
\caption{(Color online) Dependency on the lattice wavelength, with a fixed coupling-beam Rabi frequency $\Omega = 1.3\Gamma$. (a) Experimental transmission (blue) and reflection (red) spectra for several $\Delta\lambda_\lat$ and with a coupling-beam detuning
$\Delta = 2\Gamma$. (b) Corresponding simulations. (c)
Maximum reflection in the transparency window as a function of the lattice wavelength, for different detunings $\Delta$. (d) Illustration of the Bragg condition, which is fulfilled when the real part of the atomic polarizability (blue line) crosses the horizontal dashed black line representing $-\Delta\lambda_\lat/\lambda_\lat$ [Eq.~(\ref{eq.bragg})]. The red line is the imaginary part of the atomic polarizability, proportional to scattering losses.} \label{fig.RT_vs_lat}
\end{figure*}

The coupling field amplitude, parametrized by its Rabi frequency $\Omega$, is also an important parameter since the transparency increases with $\Omega$, as shown in Fig.~\ref{fig.RT_vs_EIT}(c). This increase is independent of the chosen detuning $\Delta$. The maximum reflection coefficient increases also with $\Omega$ but this time with a strong dependency on the detuning $\Delta$, as shown in Fig.~\ref{fig.RT_vs_EIT}(d). The interpretation for this behavior is the following. With a lattice wavelength such that $\Delta\lambda_\lat > 0$, the Bragg condition makes the two-level-atom band gap appear on the blue-detuned size of the atomic resonance, i.e. for $\delta \gtrsim 2$ (we recall that the atomic resonance is at $\delta \sim 2$ because of the lattice-induced light shift). As a consequence, with a large $\Delta$, like the data with $\Delta =3$, the main effect of EIT is to create a dip in the reflection band, inducing a narrow separation between the two-level-atom band gap and the EIT band gap. Then, a small $\Omega$ makes the dip smaller but does not reduce much the reflection of the EIT band gap, and that is why the maximum reflection is almost independent of $\Omega$. On the contrary, with a red-detuned coupling field (for example with $\Delta =0$), the EIT band gap is farther from the other one, and has a much lower reflectance. By looking precisely at the corresponding atomic polarizability, one can see that this is due to a higher value of Im$(\alpha)$ at the frequency where the Bragg condition (\ref{eq.bragg}) is fulfilled, inducing more losses. However, increasing $\Omega$ reduces these losses.

Therefore, this is a strong limitation for practical use of the band gap tunability: changing the coupling-beam parameters changes the atomic polarizability, which leads to more or less favorable parameters via the Bragg condition.

\subsection{Dependency on the lattice wavelength}

To study the influence of the Bragg condition, we vary the lattice wavelength \cite{footnote_depth} and record transmission and reflection spectra, for different detunings $\Delta$. First, a series of spectra obtained with the same EIT parameters is shown in Fig.~\ref{fig.RT_vs_lat}(a) with the corresponding simulations in Fig.~\ref{fig.RT_vs_lat}(b). The first notable feature is the qualitative behavior of the spectra, with the reflection band going from one side of the transmission band to the other side when the parameter $\Delta\lambda_\lat$ changes its sign. This can be easily understood by looking at the graphical representation of the Bragg condition in Fig.~\ref{fig.RT_vs_lat}(d): one can see that the frequency where the Bragg condition is fulfilled goes from one side of the maximum transparency from the other side when $\Delta\lambda_\lat$ changes its sign. Another observation is that there is a clear optimum $\Delta\lambda_\lat$ for maximizing the reflection coefficient; see the complete curves in Fig.~\ref{fig.RT_vs_lat}(c). To understand this behavior, let us first take the case with the coupling beam at resonance with the atomic transition [Figs.~\ref{fig.RT_vs_lat}(a) and \ref{fig.RT_vs_lat}(b) and $\Delta=2\Gamma$ in Figs.~\ref{fig.RT_vs_lat}(c) and \ref{fig.RT_vs_lat}(d)] and examine the limiting cases. When $\Delta\lambda_\lat\sim 0$, the Bragg condition is fulfilled where the refractive index contrast is almost zero [Eq.~(\ref{eq.bragg})], which leads obviously to an inefficient reflection. In the opposite limit, when $\Delta\lambda_\lat$ is large, the Bragg condition is fulfilled where Re$(\alpha)$ is large, but Im$(\alpha)$ is also large, inducing too much scattering loss. There is thus an optimum in between, for both signs of $\Delta\lambda_\lat$. With a nonresonant coupling field, like $\Delta =4 \Gamma$ or $\Delta = 0$ in Figs.~\ref{fig.RT_vs_lat}(c) \ref{fig.RT_vs_lat}(d), there is only one optimum lattice wavelength, for $\Delta\lambda_\lat>0$ ($\Delta\lambda_\lat<0$) with blue-detuned (red-detuned) coupling beam. This is related to the observations made in the previous paragraph: for a given $\Delta\lambda_\lat$, there is an optimum $\Delta$, and conversely, for each given $\Delta$ there is a different optimum $\Delta\lambda_\lat$. Looking very closely to a graphical representation of the Bragg condition, such as in Fig.~\ref{fig.RT_vs_lat}(d), one can always check that the difference comes from the value of Im$(\alpha)$, giving the amount of scattering losses, where the Bragg condition is fulfilled.

\section{All-optical switching} \label{sec.switch} 

Finally, to illustrate a possible application of such an atom-made tunable Bragg mirror, we demonstrate its use as an all-optical switch. This is a topic of currently high interest for the processing of optical information. Standard EIT with a disordered atomic sample can also act as an all-optical switch, since the coupling beam allows switching between transmission and absorption of the probe beam. In our case, using a periodically-ordered sample allows switching between transmission and reflection, i.e. between two output ports. Moreover, a small change in the coupling-field frequency is enough to induce switching so that full intensity modulation is not needed. In this case, the probe beam must have a fixed detuning and switching is obtained by changing the detuning $\Delta$ of the control beam, which induces a shift of the reflection band, so that for one value of $\Delta$ the probe frequency lies in the reflection band and for the other value it lies in the transmission band.

\begin{figure}[t]
\centering \includegraphics{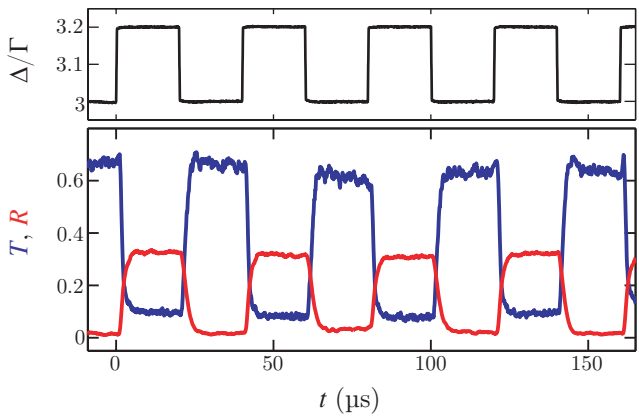} \caption{(Color online) Demonstration of a two-port all-optical switch. The detuning $\Delta$ of the coupling field serves as the control parameter (top panel). The bottom panel shows the transmission (blue) and reflection (red) coefficients as a function of time. The control-field Rabi frequency is
$\Omega = 1.5 \Gamma$, the probe detuning is $\delta = 3\Gamma$ and the lattice wavelength is such that $\Delta\lambda_\lat = 0.15$~nm.} \label{fig.AOS}
\end{figure}

We report in Fig.~\ref{fig.AOS} the result of such an experiment, with a probe beam detuning $\delta=3\Gamma$. We switch periodically the control-beam detuning  between $\Delta$ = 3$\Gamma$ and $\Delta$ = 3.2$\Gamma$ (top panel of Fig.~\ref{fig.AOS}). Following the control beam, the resulting transmission and reflection are modulated with a very good contrast, that we define by
\begin{equation}
\mathcal{C}_T = \frac{T_\mathrm{H}-T_\mathrm{L}}{T_\mathrm{H}+T_\mathrm{L}} \; , \quad \mathcal{C}_R = \frac{R_\mathrm{H}-R_\mathrm{L}}{R_\mathrm{H}+R_\mathrm{L}} \; ,
\end{equation}
where the subscripts H, L stand for the high and low levels. This leads, with the presented data, to $\mathcal{C}_T = 0.76$ and $\mathcal{C}_R = 0.88$.

Further studies are needed to better characterize the switch, in particular to determine the maximum switching rate and the minimum necessary power for the control beam. A way to achieve better performances is probably to use the four-level EIT scheme of \cite{Harris:1998}, which is known to produce giant nonlinearity, with a few photons, or ultimately a single one, being enough to make the transparency appear or disappear \cite{Yan:2001b}. This is required to enter the quantum regime, i.e., to make a quantum all-optical transistor, a key ingredient for quantum networks \cite{Kimble:2008}. Several technologies are currently investigated for realizing quantum transistors, such as plasmonic nanostructures \cite{Chang:2007}, single dye molecules embedded in crystalline matrices \cite{Hwang:2009}, ultra-high quality factor whispering-gallery-mode microresonators \cite{OShea:2011}, atoms or ions ensembles in hollow-core fibers \cite{Bajcsy:2009} or in high-finesse cavities \cite{Albert:2011,Nielsen:2011}. Since our system does not need any high-quality or microstructured mechanical elements, it might be simpler to implement.

\section{Conclusion}

We have presented in this paper a study of the photonic properties of a sample of cold atom trapped in a one-dimensional lattice under EIT conditions. In such a system, as already predicted by Petrosyan \cite{Petrosyan:2007}, EIT creates a supplementary band gap, in the transparency window, in addition to the one already present with two-level atoms \cite{Schilke:2011}. We have experimentally observed the Bragg reflection induced by this band gap and characterized its dependency with the main experimental parameters. It allowed us to put in evidence and discuss several limitations. First, the $\Lambda$ scheme necessary for EIT prevents the use of a closed transition, with an optimum transition strength, which reduces the Bragg reflection efficiency in comparison with what could be obtained with the same atomic sample by using a closed transition. In addition, the amount of scattering losses, which limit the quality of the band gaps, is at best exactly the same for the EIT band gap as for the two-level-atom band gap, and in practice slightly larger, so that the EIT band gap is of slightly lower quality. Finally, the tunability of the EIT band gap is limited by a complicated interplay between the coupling-beam parameters and the Bragg condition.

Nevertheless, it is still an interesting system, with also some advantages, like the dynamic tunability and the sharp transition between the reflection band and the transmission band. We have discussed a two-port all-optical switch as a possible application based on these properties, and we have performed a first proof-of-principle experiment. This is a promising idea that deserves further studies.

Another topic of interest is the wave propagation dynamics in this system. We have only addressed in this paper the stationary photonic properties, but it would be interesting to study pulse propagation. Both EIT and photonic band edges are known to induce slow light \cite{Dowling:1994,Scalora:1996,Hau:1999} and our system combines both ingredients. Since a short pulse is necessarily spectrally large and that, on the contrary, our system has transmission and reflection bands which are very narrow, it should induce a very large pulse distortion. This is surely not appropriate if one wants to slow down pulses without distortion, but on the contrary, a fine tuning of the parameters might allow complex and interesting pulse reshaping functions. Some proposals have recently appeared in this spirit \cite{Bariani:2008,Carusotto:2008,Bariani:2010}.

Finally, the nonlinear regime, which can be investigated by using a probe beam with a larger intensity, might also reveal interesting phenomena.


\begin{acknowledgments}
We acknowledge support from the Alexander von Humboldt foundation, the DFG and the REA (program
COSCALI, No. PIRSES-GA-2010-268717).
\end{acknowledgments}



\begin{thebibliography}{99}
\bibitem{footnote_sampling1} With $\eta = 3.5$, the r.m.s. radius of the density distribution in each period is $\sigma_z \simeq 47$~nm and we decompose each period in 40 sublayers of thickness $\delta z \simeq 9.7$~nm.
\bibitem{footnote_light_shift}
We have checked that the difference in the light-shifts undergone by the two
  involved transitions, which induces an inhomogeneous broadening of the
  two-photon resonance, was indeed negligible.
\bibitem{footnote_sampling2} For the discretizations we used the following parameters. The thickness of the sublayers in each period is $\delta z = 24$~nm, the sampling interval for the longitudinal density distribution along the lattice axis is $100~\mu$m, it is $7.5~\mu$m for the transverse distribution of the probe-beam intensity and of the atomic sample, and $0.24^\circ$ for the angle distribution that accounts for the probe beam divergence.
\bibitem{footnote_depth}
When we vary the lattice wavelength, we change the beam power accordingly in
  order to keep the same trapping depth and thus the same atom number and
  temperature.
\end{thebibliography}


\end{document}